\documentclass{ws-procs975x65}

\begin{document}

\title{LOFT: THE LARGE OBSERVATORY FOR X-RAY TIMING}

\author{ENRICO BOZZO$^*$}
\address{ISDC, University of Geneva, 
Chemin d'\'Ecogia 16, CH-1290 Versoix, Switzerland\\
$^*$E-mail: enrico.bozzo@unige.ch;\\
}

\author{PAOLO ESPOSITO}
\address{INAF--IASF Milano,
Via E. Bassini 15, I-20133 Milano, Italy\\
}

\author{PATRIZIA ROMANO}
\address{INAF--IASF Palermo,
Via U. La Malfa 153, I-90146 Palermo, Italy\\
(On the behalf of the LOFT Consortium)\\
}

\begin{abstract}
LOFT, the large observatory for X-ray timing, was selected by the European Space Agency (ESA) in February 2011 
as one of four medium size mission concepts for the Cosmic Vision program that will compete for a launch opportunity in the early 2020s.
LOFT will carry out high-time resolution (10~$\mu$s) and spectroscopic observations ($<$260~eV) of compact objects in the X-ray band 
(2-80 keV), with unprecedented throughput, thanks to its 10~m$^2$ effective area. LOFT will 
address the fundamental questions of the Cosmic Vision Theme ``Matter under extreme conditions'': 
What is the fundamental equation of state of a compact object? Does matter orbiting close to the event horizon follow the predictions 
of general relativity?
\end{abstract}


\bodymatter

\section{The LOFT mission and the on-board instruments}
\label{sec:intro} 

The LOFT mission\cite{feroci11}, selected by ESA in the context of the cosmic vision program in February 2011, is designed to exploit the diagnostics of rapid X-ray flux 
and spectral variability in order to probe directly the motion of accreting matter down to near vicinity (several Schwarzschild radii) of black holes (BHs) and neutron stars (NSs). Moreover it will determine the equation of state of 
matter at supranuclear density, through the measurement of the mass and radius of neutron star.
The LOFT payload comprises two instruments: the Large Area Detector (LAD\cite{zane12}) and the Wide Field Monitor (WFM\cite{brandt12}). 
The former is a collimated experiment (field of view, FOV, is $\sim$1~deg) with a total effective area for X-ray photons detection of about 
10~m$^2$ at 8~keV (the operating energy range is 2-80~keV). The time resolution is of $\sim$10~$\mu$s, and the 
spectral resolution $<$260~eV (FWHM EoL at 6~keV). Such a large area instrument could be designed to fit into the envelope of a medium-size mission 
thanks to the low power consumption and mass per unit area of the Silicon Drift Detectors (SDDs) and the Micro-Channel Plate collimators\cite{feroci11}. 
The WFM is a coded mask imager, observing more than 1/3 of the sky at once. It makes use of the similar SDDs to the LAD's   
(but optimized for imaging purpose), and has thus a similar time and spectral resolution. The main goal of the WFM is to detect 
source state transitions as well as  
galactic and cosmological bright events suitable for observations with the LAD. The positional accuracy 
of the WFM is of 1~arcmin. The LOFT Burst Alert System (LBAS) will broadcast the position and trigger time of bright events  
to the ground within a delay of $\sim$30~s (see also http://www.isdc.unige.ch/loft). 

Among the main science drivers of the missions is the measurement of several General Relativistic (GR) 
effects in the strong field regime (so far GR 
has been probed only through measurements in the weak field domain, i.e. $r$$\sim$10$^5$--10$^6$$r_{\rm g}$). 
As an example, the quasi-periodic oscillations (QPOs) in the X-ray flux of accreting NSs and BHs, are produced 
in the innermost regions of the accretion flow towards these objects and have been associated to 
relativistic frequencies of motion, which are much different in this regime from their Newtonian 
equivalents. By studying in detail these QPOs with the LAD it will be possible to understand 
their origin, thus getting direct access to so far untested GR effects (such as strong-field 
frame-dragging, and periastron 
precession, and the existence of an innermost stable orbit around BHs and, possibly, NSs). 
By making use of the large area and spectral capabilities of the LAD, the variable very broad 
profile of the Fe K line in galactic binaries 
and Active Galactic Nuclei will be measured at high signal-to-noise ratio, thus allowing the detection of  
Lense-Thirring nodal precession of the inner accretion disk ($\sim$$10 r_{\rm g}$) and the measurement of the mass and spin of BHs. 
Apart from detecting suitable sources and events to be re-pointed with the LAD, the WFM will also produce very valuable science by itself. 
In its large FOV, the detections of about 150 GRBs and thousands type-I X-ray bursts is expected every year 
(together with any other relatively bright transient event), and for a number of them the WFM will provide high time and 
spectral resolution data in a wide energy range (from 2 up to 80~keV). 
\begin{figure}
\begin{center}
 \includegraphics[width=1.8in]{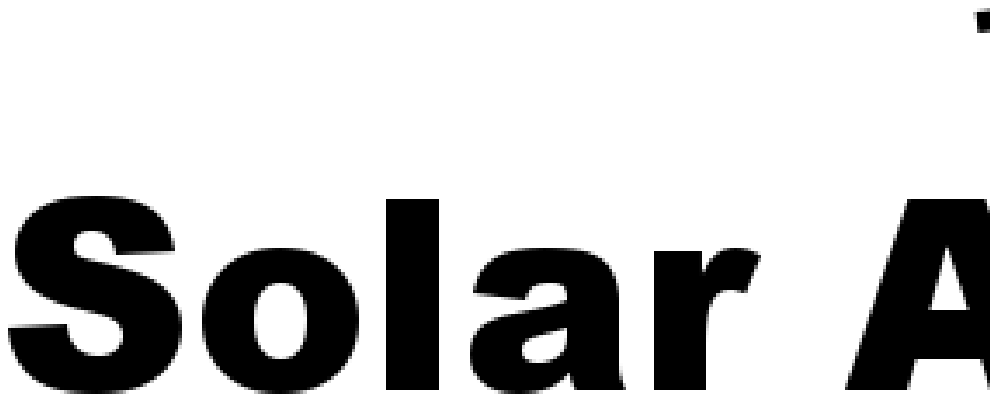}
  \includegraphics[width=1.8in]{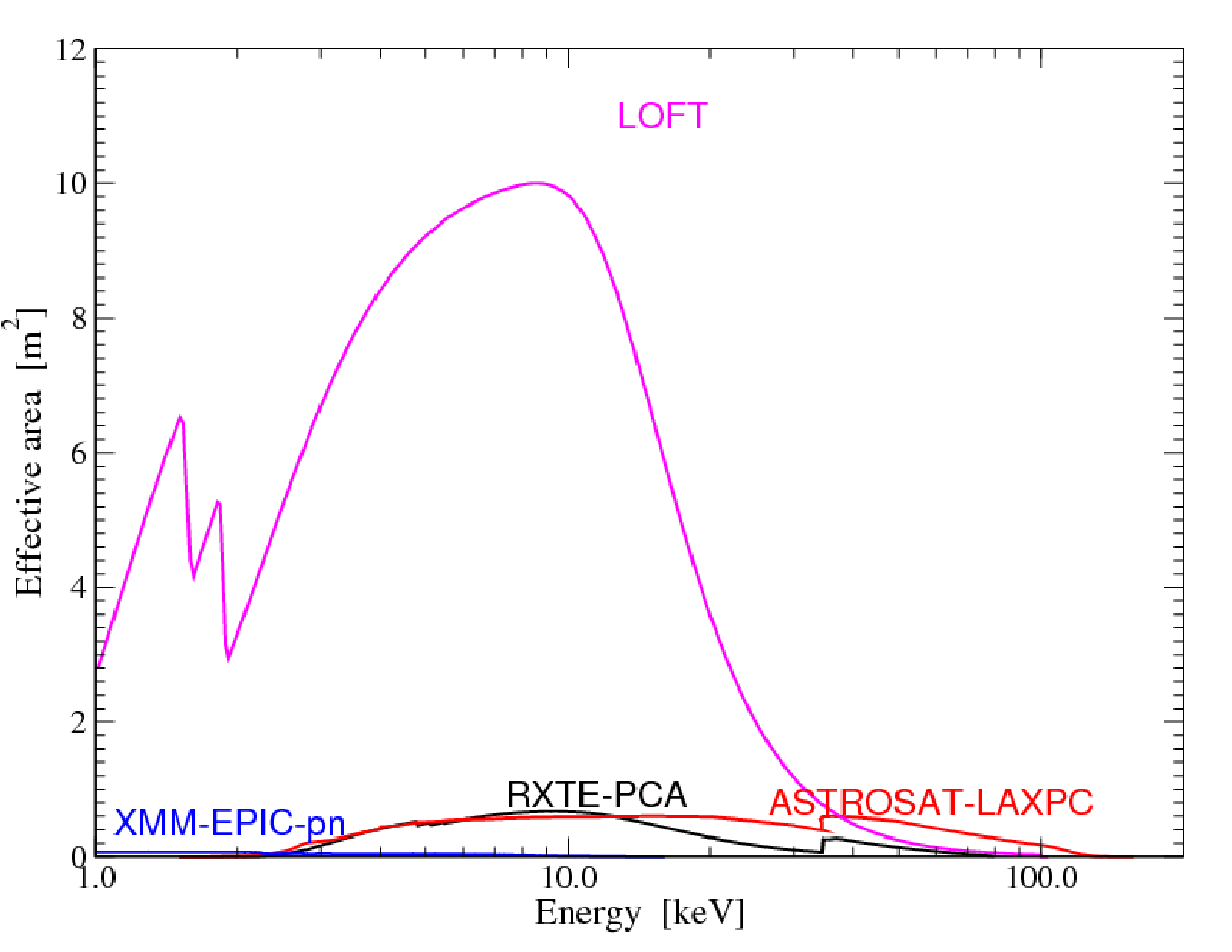}
 \caption{{\it Left}: Artist's impression of the LOFT satellite. The SDD of the LAD are placed over the 6 panels extending from 
 the spacecraft optical bench. These panels are stowed during launch and deployed once in space. The WFM is located at the top of the optical bench. 
 {\it Right}: LAD effective area as a function of the energy (other X-ray astronomy satellites are shown for comparison).} 
   \label{fig:loft}
\end{center}
\end{figure}

\end{document}